\documentclass[a4paper,11pt]{article}
\usepackage{pos}
\usepackage{subcaption}  

\title{Intelligent pixel detectors: towards a radiation hard ASIC with on-chip machine learning in 28\,nm CMOS}
\ShortTitle{Smart Pixels: Towards a rad-hard ASIC with on-chip ML in 28\,nm CMOS}

\author*[a]{Anthony Badea}
\author[b]{Alice Bean}
\author[c]{Doug Berry}
\author[d]{Jennet Dickinson}
\author[a]{Karri DiPetrillo}
\author[c,e]{Farah Fahim}
\author[c]{Lindsey Gray}
\author[c,e]{Giuseppe Di Guglielmo}
\author[h]{David Jiang}
\author[a]{Rachel Kovach-Fuentes}
\author[f]{Petar Maksimovic}
\author[g]{Corrinne Mills}
\author[h]{Mark S. Neubauer}
\author[c,g]{Benjamin Parpillon}
\author[g]{Danush Shekar}
\author[f]{Morris Swartz}
\author[c]{Chinar Syal}
\author[c,e]{Nhan Tran}
\author[g]{Jieun Yoo}

\emailAdd{anthony.badea@cern.ch}

\affiliation[a]{ The University of Chicago, Chicago, IL 60637, USA}
\affiliation[b]{ University of Kansas, Lawrence, KS 66045, USA}
\affiliation[c]{ Fermi National Accelerator Laboratory, Batavia, IL 60510, USA}
\affiliation[d]{ Cornell University, Ithaca, NY 14853, USA}
\affiliation[e]{ Northwestern University, Evanston, IL 60208, USA}
\affiliation[f]{ Johns Hopkins University, Baltimore, MD 21218, USA}
\affiliation[g]{ University of Illinois Chicago, Chicago, IL, 60607, USA}
\affiliation[h]{ University of Illinois Urbana-Champaign, Champaign, IL 61801, USA}

\abstract{
Detectors at future high energy colliders will face enormous technical challenges. Disentangling the unprecedented numbers of particles expected in each event will require highly granular silicon pixel detectors with billions of readout channels. With event rates as high as 40 MHz, these detectors will generate petabytes of data per second. To enable discovery within strict bandwidth and latency constraints, future trackers must be capable of fast, power efficient, and radiation hard data-reduction at the source. We are developing a radiation hard readout integrated circuit (ROIC) in 28nm CMOS with on-chip machine learning (ML) for future intelligent pixel detectors. We will show track parameter predictions using a neural network within a single layer of silicon and hardware tests on the first tape-outs produced with TSMC. Preliminary results indicate that reading out featurized clusters from particles above a modest momentum threshold could enable using pixel information at 40 MHz. The ICHEP presentation and proceedings are largely based on the work in Refs~\cite{Yoo:2023lxy, Parpillon:2024maz}.
}

\FullConference{42nd International Conference on High Energy Physics (ICHEP2024)\\
18-24 July 2024\\
Prague, Czech Republic\\}

\begin{document}
\maketitle

\section{Motivation}


High granularity silicon pixel detectors are essential for handling the large number of particles produced at high-energy colliders. With billions of readout channels and event rates up to 40 MHz, these detectors generate petabytes of data per second. To efficiently extract critical pixel information for physics analysis, we explore developing intelligent on-chip data reduction with a neural network (NN) approach to selectively read out pixel clusters. 

Pixel detectors, located near the interaction point, provide precise spatial measurements crucial for pattern recognition, vertexing, and momentum measurements. A charged particle passing through a pixel sensor creates a cluster of signals which can be combined with sensor location to yield precise 3D measurements. The pixel size and distance from the interaction point determine the track’s impact parameter and momentum resolution, critical measurements for physics analysis. The current ATLAS and CMS detectors contain pixel detectors with pitches of $50 \times 250-400\, \textrm{µm}^{2}$ and $100 \times 150\, \textrm{µm}^{2}$, respectively, with a thickness of $\mathcal{O}(300\,\mu \textrm{m})$. During the High Luminosity LHC era, the pixels will be reduced to roughly $50 \times 50\,\mu \textrm{m}^2$ in size and $\mathcal{O}(100\,\mu \textrm{m})$ thick \cite{CERN-LHCC-2017-021,Dominguez:1481838}.


The particle properties extracted from pixel detector data are critical for physics measurements. In high-luminosity environments, vertex information helps distinguish the primary interaction from additional proton-proton interactions occurring in the same bunch crossing (pileup). Impact parameter measurements from the pixel detector are crucial for reconstructing particles with relatively long lifetimes, such as the tau lepton, charm quark, and bottom quark. Proper identification of these particles is vital for high-priority searches and measurements, including studying the Higgs boson’s couplings and many beyond the standard model searches. 

Reading out pixel detector data is challenging due to the large data volume.
The ATLAS and CMS pixel data rates exceed bandwidth constraints at the 40 MHz collision frequency, so a hardware-based trigger reduces the event rate to less than 1 MHz. This approach often discards events that leave distinctive signatures in the pixel detector. As pixel detectors become more granular at the HL-LHC and beyond, the problem will become more extreme. Further data reduction will likely be essential to ensure important physics data is saved. To address this, we explore data reduction at the source before transmitting the pixel data. 
Further details can be found in Refs~\cite{Yoo:2023lxy, Parpillon:2024maz}.

\section{Single Cluster $p_{\mathrm{T}}$ Filtering Algorithm}

\begin{figure}[t]
  \centering
  \begin{subfigure}[b]{0.47\textwidth}
    \centering
    \includegraphics[width=\textwidth]{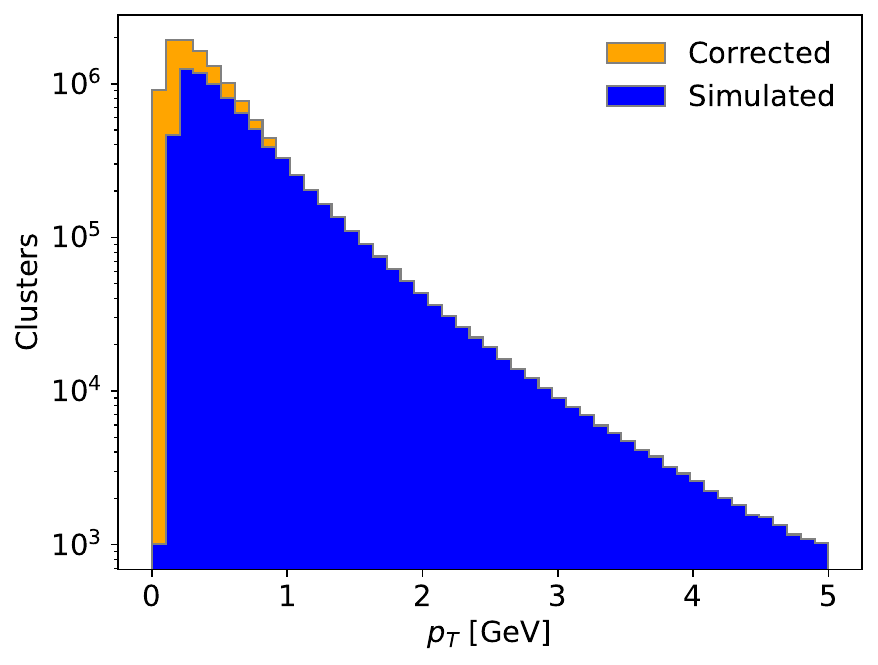}
    \caption{}
    \label{fig:pt}
  \end{subfigure}
  \hfill
  \begin{subfigure}[b]{0.47\textwidth}
    \centering
    \includegraphics[width=\textwidth]{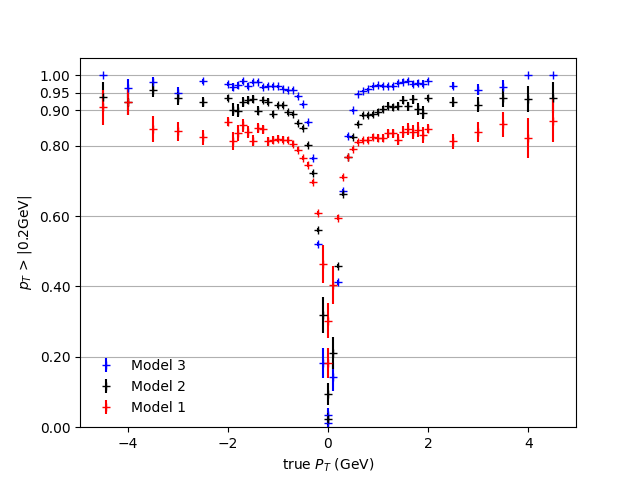}
    \caption{}
    \label{fig:curveTimingProfileYSize}
  \end{subfigure}
  \caption[]{(a) The simulated (blue) and tracking inefficiency corrected (orange) $p_{\mathrm{T}}$ distribution of tracks used to train the NN classifier. (b) Classifier acceptance as a function of $p_{\mathrm{T}}$ for three models with different input features. Positive and negative values of $p_{\mathrm{T}}$ represent the performance on clusters initiated by particles of positive and negative charge, respectively.}
  \label{fig:algorithmFigures}
\end{figure}

A NN was developed to classify single clusters arising from low vs high $p_{\mathrm{T}}$ charged particles. The studies use a simulated dataset of silicon pixel clusters produced by charged pions, with kinematic properties derived from CMS 13 TeV collision data~\cite{zenodo}. The $p_{\mathrm{T}}$ distribution starts above 100 MeV due to reconstruction limits. The distribution is corrected for CMS tracking inefficiencies as shown in Fig.~\ref{fig:pt}. The kinematics are used to seed a simulation of particles hitting a future pixel sensor with a 50\,$\mu m \times 12.5\,\mu m$ pitch, a 16$\times$16\,mm$^2$ area, a 100\,$\mu m$ thickness, and an applied bias voltage of $-$100\,V. The sensor is mounted on a 30\,mm radius cylinder in a 3.8\,T magnetic field.

The detector response was simulated using a time-sliced version of PixelAV~\cite{pixelav}, which models charge deposition by hadronic tracks, electric field maps, charge drift physics, and other key effects. PixelAV also simulates charge trapping, signal induction from trapped charge, and electronic noise, providing valuable time-evolution data for drift and induced currents in the pixel sensor. The cluster’s $x$-profile (summed over pixel columns) reflects the shape along the $x$-axis, while the $y$-profile (summed over rows) is sensitive to the particle’s incident angle and $p_{\mathrm{T}}$. The cluster shape along the $x$-axis (parallel to the magnetic field) is largely uncorrelated with $p_{\mathrm{T}}$, so just the $y$-profile was used for classification. In total, a simulated dataset of 800K clusters was used, split into 80\% for training and 20\% for testing and balanced across $p_{\mathrm{T}}$. 


Three models were developed with varying complexity to test how additional information improves $p_{\mathrm{T}}$ discrimination. The models were built to predict whether a cluster arose from three classes of charged particles: $p_{\mathrm{T}}>200$ MeV, and $p_{\mathrm{T}}<200$ MeV for positively and negatively charged particles. Two background classes were used because oppositely charged particles in a magnetic field were found to leave different cluster shapes. The threshold of 200 MeV was chosen to ensure flat signal efficiency for tracks with $p_{\mathrm{T}}>2$ GeV, a threshold important for physics analysis based on min-bias studies~\cite{Ryd:2020ear}. Model 1 is a two layer dense nueral network (DNN) with only 2 inputs , the cluster position ($y_{0}$) and size $y$-size. It correctly selects roughly 85\% of tracks with $p_T>2$ GeV. Model 2 uses the full cluster $y$-profile and the same architecture as Model 1, improving accuracy to 93.3\% for $p_{\mathrm{T}}>2$ GeV, while remaining compact enough for hardware implementation. Model 3 is a Convolutional Neural Network (CNN) that operates on 20 time-stamps of the $y$-profile and achieves the highest overall accuracy with an additional 4\% gain in signal efficiency. The 3 model efficiencies versus track $p_{\mathrm{T}}$ are shown in Fig.~\ref{fig:curveTimingProfileYSize}. In spite of Model 3's improved performance, the complexity of time-sliced data extraction poses challenges for chip design, so Model 2 was chosen for hardware implementation. The bandwidth saving using Model 2 was estimated to be in the range $57 - 76\%$.


\section{On-chip Implementation}

\begin{figure}[t]
  \centering
  \begin{subfigure}[b]{0.47\textwidth}
    \centering
    \includegraphics[width=\textwidth]{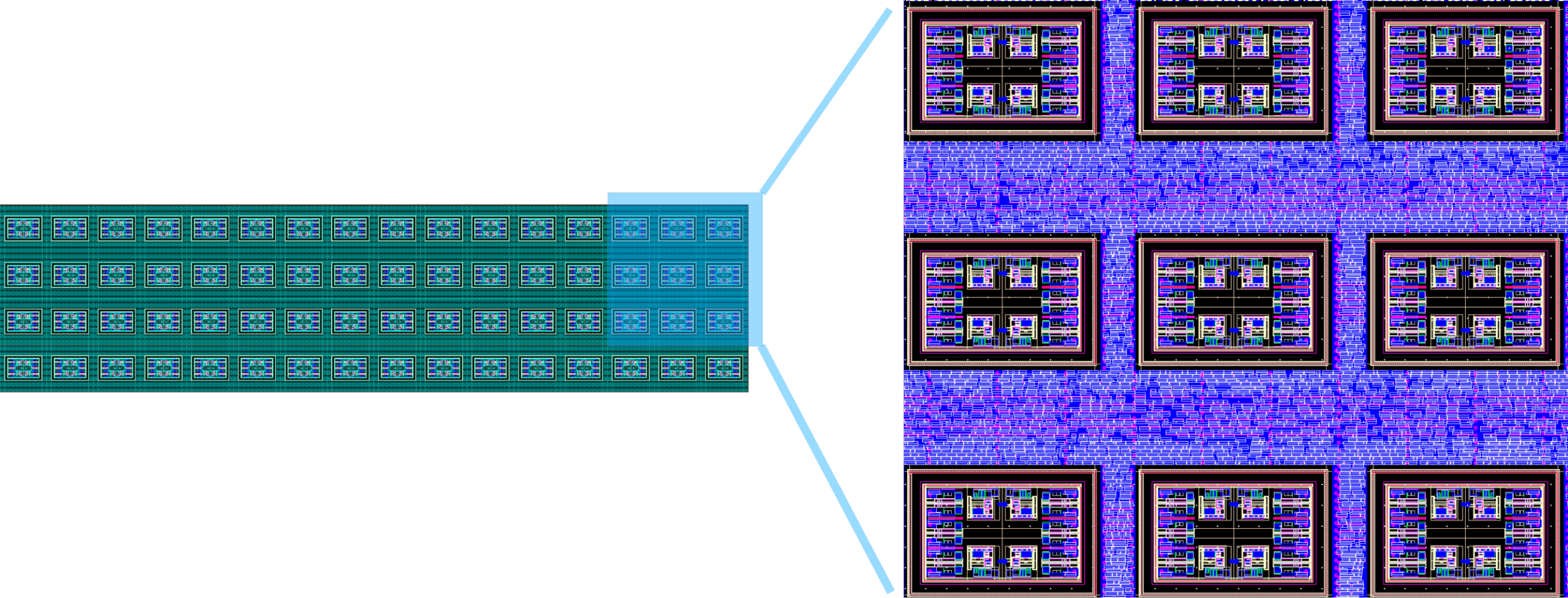}
    \caption{}
    \label{fig:chip_tapeout}
  \end{subfigure}
  \hfill
  \begin{subfigure}[b]{0.47\textwidth}
    \centering
    \includegraphics[width=\textwidth]{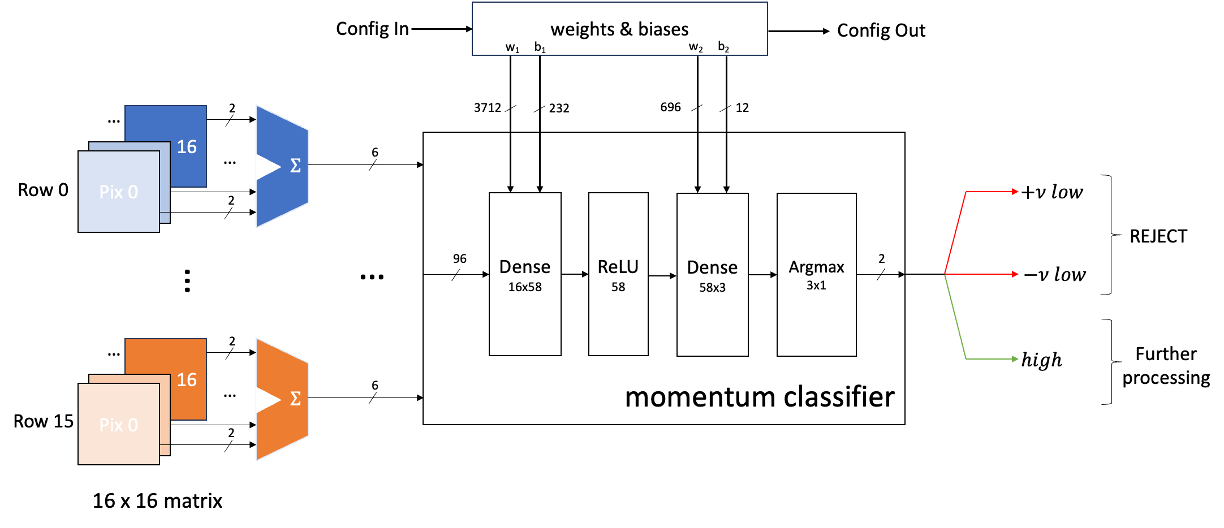}
    \caption{}
    \label{fig:SuperPixelImplementation}
  \end{subfigure}
  \caption[]{(a) 2x2 pixel analog islands (within black boxes) surrounded by digital logic within deep neural network (DNN) and test interface (purple space). Chip taped-out as super pixels (16x4), corresponding to 32x8 physical pixels. (b) Data flow through the digital implementation of the algorithm from the summed ADC bits (on the left) through the neural network and the final classification layer. At the top of the diagram we illustrate the reconfigurability of the weights and biases in the algorithm stored in memory.}
  \label{fig:chipImplementation}
\end{figure}

A smart pixel prototype ROIC was designed as a collections of 2x2 pixels consisting of analog islands surrounded by digital logic where the filtering NN sits. Each ROIC pixel is 25x25 $\mu$m$^{2}$. Two $32x8$ arrays of pixels were taped-out on a 1.6\, mm$^{2}$ ASIC chip. A TSMC CMOS 28\,nm bulk process was used. The design is shown in Figure~\ref{fig:chip_tapeout} and a brief summary of key aspects is provided below.

The analog islands are designed as follows. The charge collected at the sensor's electrode is integrated, amplified, and converted to voltage using a charge-sensitive preamplifier. An AC-coupled 2-bit flash-type ADC digitizes the signal. Due to the thermometric nature of the flash ADC in our design, analog-to-digital conversion begins as soon as the integrated charge output is above the first threshold, and continues until the signal reaches its maximum value or the time for conversion runs out.
Further details are discussed in \cite{ISCAS2023}. 


The NN was translated to a hardware implementation using \texttt{hls4ml}~\cite{fahim2021hls4ml}, an open-source Python framework that facilitates the co-design of ML algorithms for hardware deployment, supporting models from quantized models from QKeras and other formats~\cite{coelho2021automatic}. We fine-tuned the numerical precision and the hardware parallelism to optimize area, performance, and power consumption. The conversion process started with the quantized model of the classifier, which \texttt{hls4ml} translated into HLS-ready C++ code for Siemens Catapult HLS~\cite{catapult-hls} that generates a hardware description at the register-transfer level (RTL) suitable for the ASIC flow. We chose to fully parallelize the hardware logic to minimize the latency of the neural network, integrating the HLS-generated RTL design with system registers and data movers for efficient operation. A sketch of the data flow through the digital implementation is shown in Figure~\ref{fig:SuperPixelImplementation}. The NN consumes around 300\,$\mu$W per $32x8$ pixel array, assuming an estimated occupancy of 1 hit per mm$^{2}$. The overall power consumption per pixel, including analog and digital functions, is 6\,$\mu$W, resulting in approximately 1 W/cm$^{2}$, within the permissible limits of the HL-LHC experiments~\cite{Garcia-Sciveres:2663161}.

\section{Timing Violation Test with a Python-Driven Test Stand}

\begin{figure}[t]
  \centering
  \begin{subfigure}[b]{0.47\textwidth}
    \centering
    \includegraphics[width=\textwidth]{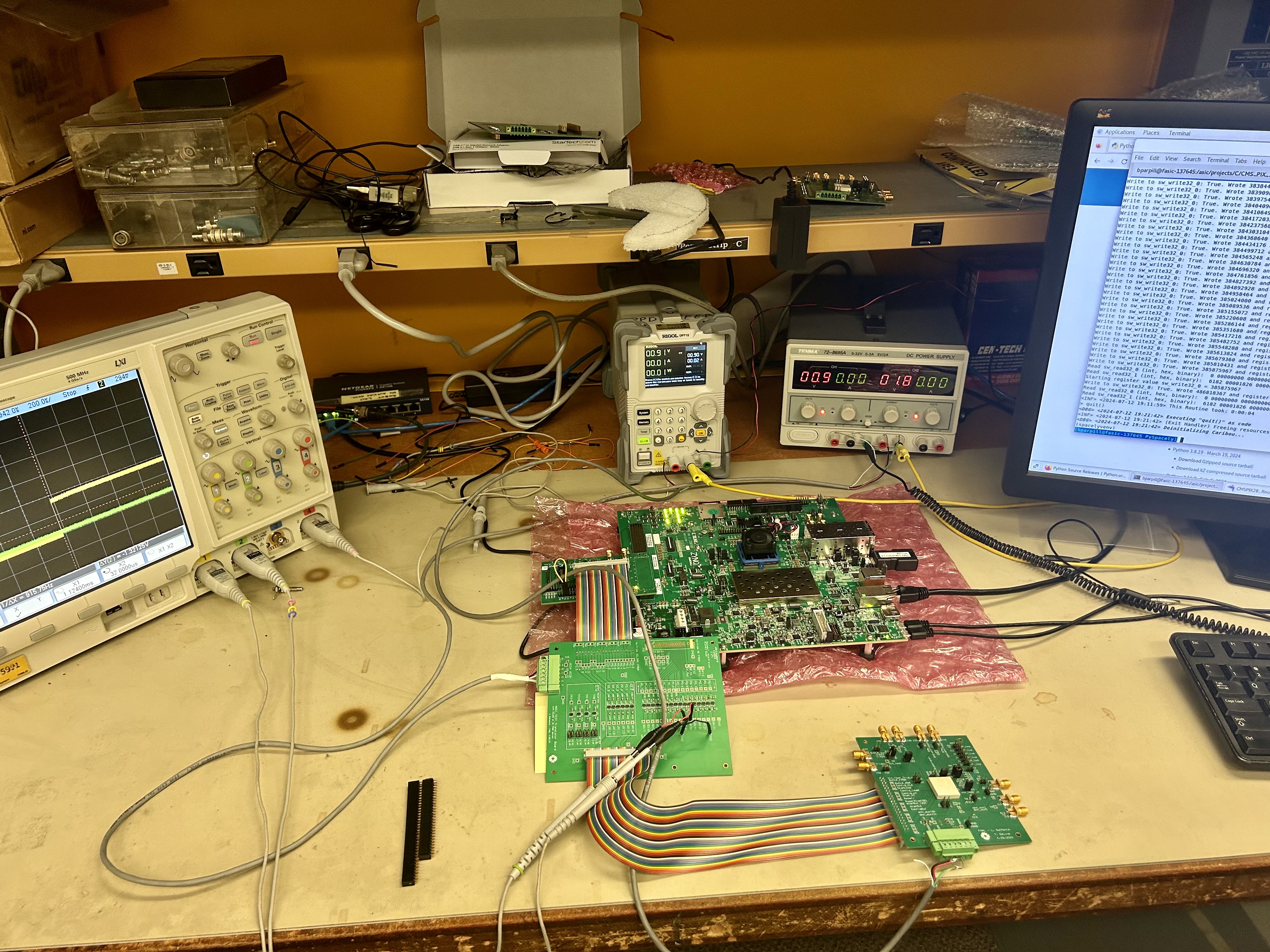}
    \caption{}
    \label{fig:fnal_lab_setup}
  \end{subfigure}
  \hfill
  \begin{subfigure}[b]{0.47\textwidth}
    \centering
    \includegraphics[width=\textwidth]{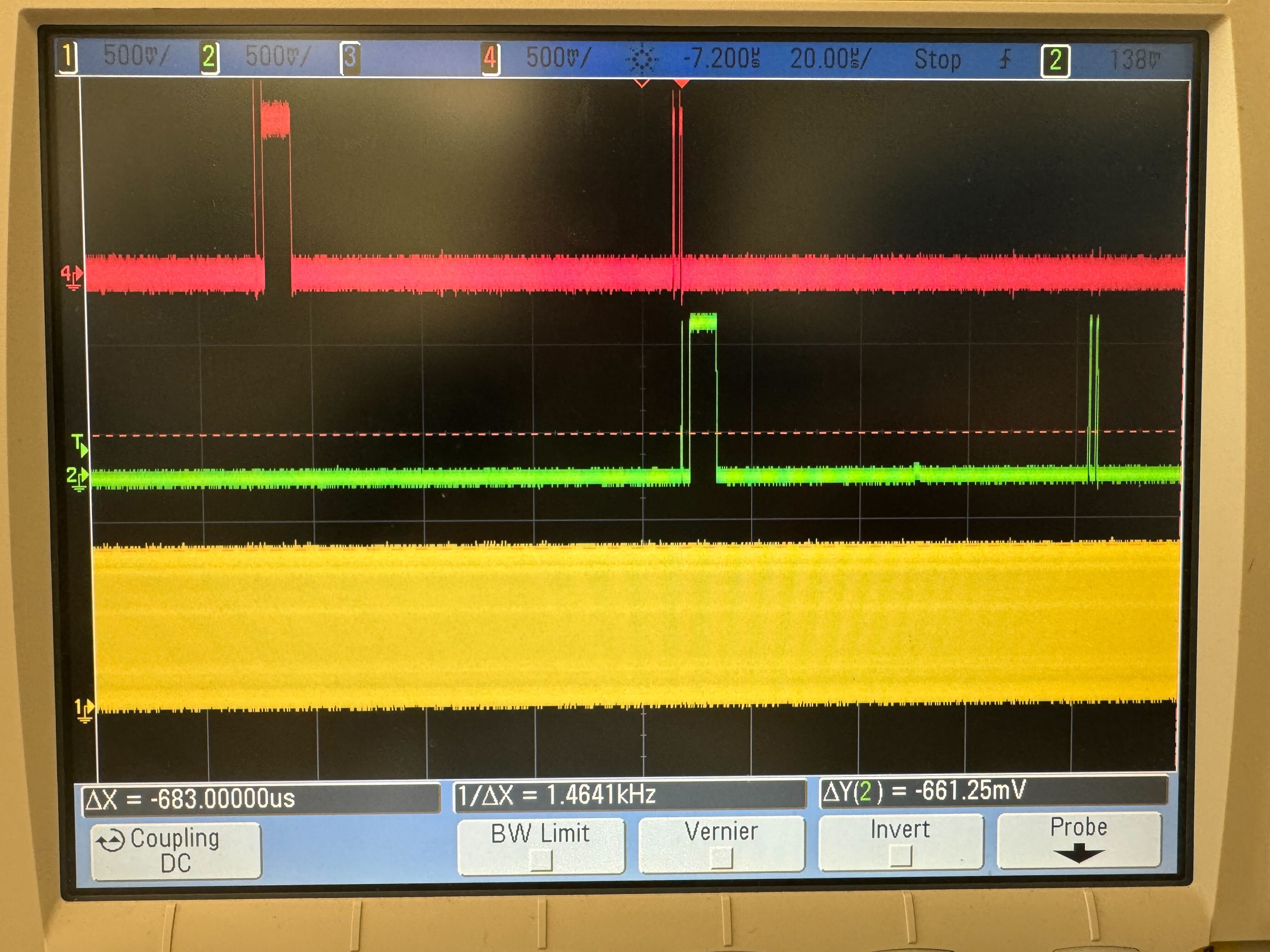}
    \caption{}
    \label{fig:timing_volation_check}
  \end{subfigure}
  \caption[]{(a) The test stand consisting of a Linux machine running Spacely (monitor), a Xilinx ZCU102 System on Chip (SoC) FPGA (center), a mezzanine and level shifter board (center left), and the device under test (DUT) (center right). (b) A snapshot from the oscilloscope used to monitor the timing violation test. A 10 MHz clock is transmitted to the DUT (yellow). Input pulses (red) are sent to the DUT. The DUT shifts those patterns through 768 registers, one register every clock cycle. The pattern is read back on the scope via an output register (green). The expected time delay between when the pattern is inputted and when it appears at the output register is seen.}
  \label{fig:fnalTestStand}
\end{figure}

A test stand was built at Fermilab to perform python-driven ASIC testing by leveraging the open-source Spacely workflow~\cite{Quinn:2024xhl}. The setup is shown in Figure~\ref{fig:fnal_lab_setup}, consisting of a PC running Spacely, a Xilinx ZCU102 System on Chip (SoC) FPGA, a mezzanine and level shifter board, and the device under test (DUT). Python routines are executed on the Linux machine which sends commands at low rate to the SoC’s microprocessor. The SoC buffers the commands and communicates with custom firmware on the FPGA’s programmable logic (PL) via a CERN/BNL-created interface~\cite{Vanat:2703500}. The PL then sends high-rate commands through the mezzanine and level shifter board to the DUT, which is a printed circuit board bonded to the ROIC. The communication pipeline is bidirectional, allowing data to be read back from the DUT and analyzed on the PC. Breakout pins on the level shifter and DUT are utilized to inspect the signals on an oscilloscope.

The test stand was used to carry out several first tests of the ROIC, including a time violation test. In the test, a 10 MHz clock was generated by the FPGA and transmitted to the DUT. Then, multiple bits were sent to the DUT. The DUT was instructed to serially shift the bits through 768 registers, moving one register every clock cycle. At the end of 768 clock cycles, the output register was expected to show the input pattern. The output register was monitored on an oscilloscope while the test was performed. A snapshot of the test is shown in Figure~\ref{fig:timing_volation_check}. The input patterns, output register, and clock are shown in red, green, and yellow, respectively. We observe the expected 768 clock cycles between each of the input patterns and the resulting output pattern. This validates that up to 10 MHz the chip is free of timing violations. Further tests are needed with high statistics pulsing and at higher rates up to the desired 40 MHz frequency. The core functionality of the test stand, including new python routines, firmware blocks, and auxiliary pulsing hardware, is currently being deployed to perform those tests.

\section{Conclusions}

Silicon pixel sensors are critical for high-energy physics experiments, but fully utilizing them in high-rate environments such as the HL-LHC requires innovative approaches. This study explores on-sensor ML for filtering low-momentum tracks to reduce data rates and make pixel information usable in online trigger systems. We successfully trained a NN to differentiate high vs. low $p_{\mathrm{T}}$ particles traversing a small-pitch pixel sensor in a magnetic field. Early results demonstrated moderate bandwidth reduction, motivating its integration in an on-chip implementation. The ROIC was designed in CMOS 28\,nm bulk technology with analog pixel islands and surrounding digital logic to house the neural network. 
A Python-driven test stand was built, and initial timing checks were successful. Future tests aim to characterize the analog components and evaluate the NN’s performance. These results highlight the potential of on-chip ML to enhance high-rate particle physics experiments, motivating further studies into innovative data reduction methods.

\section{Acknowledgements}

Thank you to all those included in the acknowledgements of Ref~\cite{Yoo:2023lxy} as well as the ICHEP organizers. AB is supported by Schmidt Sciences, LLC and utilized the UChicago AI+Science computing cluster for part of this work.

\bibliography{bib}

\end{document}